\documentclass[12pt,a4paper]{article}

\usepackage[utf8]{inputenc}
\usepackage{fullpage}
\usepackage[english]{babel}
\usepackage{amsmath}
\usepackage{amssymb}
\usepackage{amsthm}
\usepackage[hidelinks]{hyperref}
\usepackage{commath}
\usepackage[T1]{fontenc}
\usepackage{tikz}
\usepackage{appendix}

\title{Elementary Microeconomics of the Talmudic Rule}
\author{Anton Salikhmetov}

\newcommand{\dr}{\%\Delta}
\newcommand{\bp}{\text{\textpertenthousand}}

\begin{document}
\maketitle

\begin{abstract}
This paper takes a look at the Talmudic rule aka the $1/N$ rule aka the uniform investment strategy from the viewpoint of elementary microeconomics.
Specifically, we derive the cardinal utility function for a Talmud-obeying agent which happens to have the Cobb--Douglas form.
Further, we investigate individual supply and demand due to rebalancing and compare them to market depth of an exchange.
Finally, we discuss how operating as a liquidity provider can benefit the Talmud-obeying agent with every exchange transaction in terms of the identified utility function.
\end{abstract}

\section{Deriving the cardinal utility function}

The Talmudic rule\footnote{
``A person should always divide his money into three; he should bury one-third in the ground,
and invest one-third in business, and keep one-third in his possession'' (Bava Metzia 42a).
} requires one to store equal amount of goods and money with respect to the market price of the goods at any given time.
Let us assume that an agent has initially bought a quantity $q_0$ of goods at the price $p_0$ and is left with $m_0$ units of money.
Then we can derive the cardinal utility function $u(m, q)$ of that agent measured in the units of money.
First, the initial configuration requires the budget $2 m_0 = 2 p_0 q_0$, thus we have $u(m_0, q_0) = 2 p_0 q_0$.
Second, having $k m$ of money and $k q$ of goods should be exactly $k$ times more preferable than having $m$ of money and $q$ of goods.
Finally, notice that the Talmud-obeying agent should always be ready to exchange $|\Delta q|$ of goods for $|\Delta m|$ of money (and vice versa) as long as the exchange rate is equal to the ratio between the resulting amounts of money and goods in their possession:
\begin{align*}
\frac{\Delta m}{\Delta q} &= -\frac{m + \Delta m}{q + \Delta q}; \\
\frac{dm}{dq} &= \lim_{\Delta q \rightarrow 0} \frac{\Delta m}{\Delta q} = -\frac{m}{q}; \\
\frac{dm}{m} + \frac{dq}{q} &= 0; \\
\ln m + \ln q &= \ln C; \\
m q &= C.
\end{align*}
The latter means that $u(m, q)$ remains constant along hyperbolas $m q = C$.
Thus we have $u(m, q) = f(m q)$ for some function $f(t)$.
As noticed earlier, $u(km, kq) = k u(m, q)$, hence $f(t) = \sqrt t f(1)$ and $u(m, q) = f(1) \sqrt{m q}$.
Taking into account $u(p_0 q_0, q_0) = 2 p_0 q_0$, we conclude with $f(1) = 2 \sqrt{p_0}$ and $u(m, q) = 2 \sqrt{p_0 m q}$.
Note that the latter utility function has the Cobb--Douglas form $A m^\alpha q^{1-\alpha}$ with efficiency $A = 2 \sqrt{p_0}$ and elasticity $\alpha = 1/2$.
The resulting indifference map is illustrated in Figure~\ref{indiff}.
See also Appendix~\ref{weighted}.

\begin{figure}
\begin{minipage}{0.5\textwidth}
\caption{Indifference map.}
\label{indiff}
\begin{center}
\begin{tikzpicture}[scale=2]
\path (-0.5,-0.3) rectangle (3,2.8);
\draw [help lines,smooth,domain=0:2] plot (\x,\x);
\draw [fill] (0,0.5) circle (0.5pt);
\draw [fill] (0,1) circle (0.5pt);
\draw [fill] (0,1.5) circle (0.5pt);
\draw [fill] (0.5,0) circle (0.5pt);
\draw [fill] (1,0) circle (0.5pt);
\draw [fill] (1.5,0) circle (0.5pt);
\draw [fill] (0.5,0.5) circle (0.5pt);
\draw [fill] (1,1) circle (0.5pt);
\draw [fill] (1.5,1.5) circle (0.5pt);
\draw [fill] (0,0) circle (0.5pt);
\node [below left] at (0,0) {$0$};
\node [left] at (0,2.5) {$m$};
\node [below] at (2.5,0) {$q$};
\draw [<->] (0,2.5) -- (0,0) -- (2.5,0);
\draw [thick,smooth,domain=0.5:2] plot (\x,{1/\x});
\draw [thick,smooth,scale=0.5,domain=0.25:4] plot (\x,{1/\x});
\draw [thick,smooth,scale=1.5,domain=0.75:4/3] plot (\x,{1/\x});
\node [right] at (2,0.125) {$u = 2 m_0$};
\node [right] at (2,0.5) {$u = 4 m_0$};
\node [right] at (2,1.125) {$u = 6 m_0$};
\node [above right] at (2,2) {$m = p_0 q$};
\draw [dashed] (0,0.5) -- (0.5,0.5) -- (0.5,0);
\draw [dashed] (0,1) -- (1,1) -- (1,0);
\draw [dashed] (0,1.5) -- (1.5,1.5) -- (1.5,0);
\node [left] at (0,0.5) {$m_0$};
\node [left] at (0,1) {$2m_0$};
\node [left] at (0,1.5) {$3m_0$};
\node [below] at (0.5,0) {$q_0$};
\node [below] at (1,0) {$2q_0$};
\node [below] at (1.5,0) {$3q_0$};
\end{tikzpicture}
\end{center}
\end{minipage}%
\begin{minipage}{0.5\textwidth}
\caption{Supply and demand.}
\label{snd}
\begin{center}
\begin{tikzpicture}[scale=2]
\path (-0.5,-0.3) rectangle (3,2.8);
\draw [fill] (0,0) circle (0.5pt);
\draw [fill] (0,0.5) circle (0.5pt);
\draw [fill] (1.5,0) circle (0.5pt);
\draw [dashed] (1.5,2) -- (1.5,0);
\node [below left] at (0,0) {$0$};
\node [left] at (0,2.5) {$p$};
\node [below] at (2.5,0) {$q$};
\node [left] at (0,0.5) {$p_0$};
\node [below] at (1.5,0) {$q_0$};
\draw [<->] (0,2.5) -- (0,0) -- (2.5,0);
\draw [thick,smooth,domain=0:0.75] plot (\x,{1/(2*(1-\x/1.5)*(1-\x/1.5))}) node [left] {$p_s$};
\draw [thick,smooth,domain=0:2] plot (\x,{1/(2*(1+\x/1.5)*(1+\x/1.5))}) node [above] {$p_d$};
\end{tikzpicture}
\end{center}
\end{minipage}
\end{figure}

\section{Supply and demand due to rebalancing}

Note that the indifference curve $m q = m_0 q_0$ is defined in terms of amounts $m$ and $q$ that are in \textit{possession} of our Talmud-obeying agent, so it cannot be used directly to find the supply and demand curves.
Instead, we should consider supplied $q_s$ and demanded $q_d$ quantities due to rebalancing required by the Talmudic rule when the price changes.

Specifically, when the price goes up, our agent is to sell some $q_s = q_0 - q$ of their $q_0$ units of goods, resulting in $q$ units left in their possession.
Note that since $q > 0$, we have $q_s < q_0$.
Conversely, when the price goes down, the agent buys $q_d = q - q_0$ units of goods in addition to their $q_0$ units and is left with the total of $q$ units.
As the Talmudic rule implies $m = p q$ at any given price $p$, we can rewrite the indifference curve as $p q^2 = p_0 q_0^2$ and use it in order to obtain the following supply and demand curves:
$$
p_s = \frac{p_0 q_0^2}{(q_0 - q_s)^2}, \quad
p_d = \frac{p_0 q_0^2}{(q_0 + q_d)^2}.
$$
These supply and demand curves are illustrated in Figure \ref{snd}.
Notice that due to limited resources $q_0$ and $m_0$ in possession of our Talmud-obeying agent, the supply curve has an asymptote $q = q_0$, whereas the total area below the demand curve is equal to $m_0$:
$$
\int_0^\infty \frac{p_0 q_0^2}{(q_0 + q_d)^2}\,dq_d = p_0 q_0.
$$

Figure \ref{snd} might look familiar.
Indeed, it resembles market depth that is provided by exchanges, when the current price is equal to $p_0$.
Notice that near $p = p_0$, our supply and demand curves behave linearly with the same absolute slope.
The similar behavior can be noticed about market depth at relatively stable conditions near the current price.
The latter observation about market depth has a rather curious consequence.
Namely, we can estimate the budget $u_0 = 2 p_0 q_0$ that is needed for a Talmud-obeying agent in order to provide as much liquidity near the current price as the whole exchange does:
\begin{align*}
\frac{dp_s}{dq_s} &= -\frac{dp_d}{dq_d} = \frac{2p_0}{q_0}\ \text{at $q_s = q_d = 0$}; \\
u_0 &= 2 p_0 q_0 \approx 4p_0^2 \cdot \left|\frac{\Delta q}{\Delta p}\right|.
\end{align*}

\section{Operating as a market maker}

In the previous sections, we applied the Talmudic rule to infinitesimal changes in price.
However, situation is slightly different for any finite non-zero change in price.
As discussed in \cite[Section 2]{zigzag}, the Talmudic rule then implies that the geometric mean of the quantity of goods and the amount of money in possession always increases:
\begin{equation}
\label{gm}
\frac{u_1}{u_0} = \frac{u(p_1 q_1, q_1)}{u(p_0 q_0, q_0)} = \frac{\sqrt{m_1 q_1}}{\sqrt{m_0 q_0}} = \frac{p_0 + p_1}{2 \sqrt{p_0 p_1}} > 1.
\end{equation}

Let us denote the relative differences in utility and price, respectively, as follows:
$$
\dr u = 2 \cdot \frac{|u_0 - u_1|}{u_0 + u_1}, \quad \dr p = 2 \cdot \frac{|p_0 - p_1|}{p_0 + p_1}.
$$
Then, we can use (\ref{gm}) to calculate the relative growth in utility $\dr u$ after the transaction that corresponds to a relative change in price $\dr p$, assuming $p_1 > p_0$:
\begin{align*}
\frac{u_1}{u_0} &= \frac{2 + \dr u}{2 - \dr u}, \quad
\frac{p_1}{p_0} = \frac{2 + \dr p}{2 - \dr p}; \\
\frac{u_1}{u_0} &= \frac{1}{2} \left(\sqrt{\frac{p_1}{p_0}} + \sqrt{\frac{p_0}{p_1}}\right); \\
\dr u &= \frac{16 - 8 \sqrt{4 - (\dr p)^2}}{(\dr p)^2} - 2; \\
\dr u &= \frac{(\dr p)^2}{8} + \frac{(\dr p)^4}{64} + O\left((\dr p)^6\right)\ \text{at $\dr p = 0$}.
\end{align*}
For example, $\dr p = 2\%$ results in $\dr u \approx 0.5\bp$.
With such a transaction once a day on average and assuming negligible transaction fees, the annual interest rate in terms of our Talmudic utility is about $1.84\%$.
As one would expect, the market maker has to face a trade-off between the frequency of transactions and the utility growth per transaction.
In turn, \cite{zigzag} suggests that to find the optimal values of $\dr p$ may not be a trivial problem.

\appendixtitleon
\begin{appendices}
\section{The weighted Talmudic rule}
\label{weighted}

This section considers an arbitrarily weighted version of the Talmudic rule.
Specifically, let $\alpha, \beta \in (0, 1)$ and $\alpha + \beta = 1$.
Then the weighted Talmudic rule would require one to store portion $\alpha$ of value in money and $\beta$ in goods with respect to the market price.

Let $u(m, q)$ be the corresponding cardinal utility function.
If $u_0 = u(m_0, q_0)$, then we have $m_0 = \alpha u_0$ and $p_0 q_0 = \beta u_0$.
Note $u(km, kq) = k u(m, q)$.
The rest is straightforward:
\begin{align*}
p &= \frac{\beta}{\alpha} \cdot \frac{m}{q}; \\
\frac{\Delta m}{\Delta q} &= -\frac{\beta}{\alpha} \cdot \frac{m + \Delta m}{q + \Delta q}; \\
\frac{dm}{dq} &= \lim_{\Delta q \rightarrow 0} \frac{\Delta m}{\Delta q} = -\frac{\beta}{\alpha} \cdot \frac{m}{q}; \\
\alpha \cdot \frac{dm}{m} + \beta \cdot \frac{dq}{q} &= 0; \\
\alpha \ln m + \beta \ln q &= \ln C; \\
m^\alpha q^\beta &= C; \\
u(m, q) &= f(m^\alpha q^\beta); \\
f(t) &= f(t^\alpha t^\beta) = t f(1); \\
f(1) &= \frac{u_0}{m_0^\alpha q_0^\beta} = \frac{p_0^\beta}{\alpha^\alpha \beta^\beta}; \\
u(m, q) &= \frac{p_0^\beta}{\alpha^\alpha \beta^\beta} \cdot m^\alpha q^\beta.
\end{align*}
Further, we use the indifference curve $m^\alpha q^\beta = m_0^\alpha q_0^\beta$ with $q_s = q_0 - q$ and $q_d = q - q_0$ in order to obtain the supply and demand curves due to weighted rebalancing:
\begin{align*}
p_s &= \frac{p_0 q_0^{1/\alpha}}{(q_0 - q_s)^{1/\alpha}}; &
p_d &= \frac{p_0 q_0^{1/\alpha}}{(q_0 + q_d)^{1/\alpha}}; \\
q_s &< q_0; &
\int_0^\infty p_d \, dq_d &= \frac{\alpha}{\beta} \cdot p_0 q_0 = m_0; \\
\frac{dp_s}{dq_s} &= \frac{p_0}{\alpha q_0}\ \text{at $q_s = 0$}; &
\frac{dp_d}{dq_d} &= -\frac{p_0}{\alpha q_0}\ \text{at $q_d = 0$}.
\end{align*}

For a finite change in price from $p_0$ to $p_1$, we have the ratio between $u_1$ and $u_0$ equal to the weighted arithmetic mean of the prices divided by their weighted geometric mean:
$$
\frac{u_1}{u_0} = \frac{\alpha p_0 + \beta p_1}{p_0^\alpha p_1^\beta}\ \Rightarrow\ \dr u = \frac{\alpha \beta}{2} \cdot (\dr p)^2 + O\left((\dr p)^3\right)\ \text{at $\dr p = 0$}.
$$
Let us notice that the case of $\alpha = \beta = 1/2$ is special in the following two aspects.
First, it is the only proportion in which $u_1/u_0$ ratio is invariant to the direction of change in price, the latter being crucial for unpredictable markets.
Second, $\alpha = \beta = 1/2$ maximizes $\dr u$ for small changes in prices which can be very useful for liquidity providers.
\end{appendices}
\end{document}